\documentclass[a4paper,aps,prd,10pt,preprintnumbers,showpacs,onecolumn,superscriptaddress,nofootinbib,amsmath,amssymb]{revtex4-1}
\usepackage{amssymb}
\usepackage{amsfonts}
\usepackage{amsmath}
\usepackage{graphicx}
\usepackage{dcolumn}
\usepackage{bm}
\usepackage{makeidx}
\usepackage[latin1]{inputenc}
\usepackage[english, english]{babel}
\usepackage{color}

\usepackage{graphicx}
\def\be{\begin{equation}}
 \def\ee{\end{equation}}
 \def\bea{\begin{eqnarray}}
 \def\eea{\end{eqnarray}}

\def\aa{\widetilde{\alpha}}

\makeindex

\begin{document}

\title{Perturbative and nonperturbative 
fermionic quasinormal modes 
of Einstein--Gauss--Bonnet--AdS black holes}
\author{P. A. Gonz\'{a}lez}
\email{pablo.gonzalez@udp.cl}
\affiliation{Facultad de Ingenier\'{\i}a y Ciencias, Universidad Diego Portales, Avenida Ej\'{e}%
rcito Libertador 441, Casilla 298-V, Santiago, Chile.}

\author{Yerko V\'{a}squez}
\email{yvasquez@userena.cl}
\affiliation{Departamento de F\'isica y Astronom\'ia, Facultad de Ciencias, Universidad de La Serena,\\
Avenida Cisternas 1200, La Serena, Chile.}

\author{Ruth Noem\'i Villalobos}
\email{rvillalobos@userena.cl}
\affiliation{Departamento de F\'isica y Astronom\'ia, Facultad de Ciencias, Universidad de La Serena,\\
Avenida Cisternas 1200, La Serena, Chile.}

\date{\today}

\begin{abstract}

In this work, we present the quasinormal modes of a fermionic field in the background of Gauss--Bonnet--AdS black holes. We find exact solutions for $D=5$ at the fixed value $\alpha=R^2/2$ of the Gauss--Bonnet coupling constant, with $R$ denoting the AdS radius, and we find numerical solutions for some range of values of the coupling constant $\alpha$ and $D=5, 6$. Mainly, we find two branches of quasinormal frequencies, a branch perturbative in the Gauss--Bonnet coupling constant $\alpha$, and another branch nonperturbative in $\alpha$. The phenomena of nonperturbative modes, which seem to be quite general in theories with higher curvature corrections, have been obtained in the spectrum of gravitational field perturbations and scalar field perturbations in previous works. We show that it also arises for fermionic field perturbations and therefore seems to be independent of the spin of the field under consideration. However, in contrast to gravitational and scalar field perturbations, where the nonperturbative modes are purely imaginary, we find that for fermionic field perturbations the nonperturbative modes acquire a real part. We find that the imaginary part of the quasinormal frequencies is always negative in both branches; therefore, the spherical Gauss--Bonnet--AdS black holes are stable against fermionic field perturbations.

\end{abstract}

\maketitle


\tableofcontents


\newpage
\section{Introduction}

Quasinormal modes (QNMs) and quasinormal frequencies (QNFs) have been a subject of study for a long time 
\cite{Regge:1957td,Zerilli:1971wd, Kokkotas:1999bd, Nollert:1999ji, Konoplya:2011qq} and  have recently acquired great interest due  
to the detection 
of gravitational waves \cite{Abbott:2016blz}.
Despite the detected signal being consistent with the Einstein gravity \cite{TheLIGOScientific:2016src}, there are great uncertainties in mass and angular momenta of the ringing black hole, which leaves open possibilities 
for alternative theories of gravity
\cite{Konoplya:2016pmh}. In this sense, higher order curvature corrections to the Einstein gravity, such as 
Gauss--Bonnet terms, are an interesting alternative, mainly because they appear at the low-energy limit of string theory. QNMs are important in the analysis of the classical stability of black holes against probe matter fields. On the other hand,  QNMs are important in the context of the AdS/CFT correspondence, and the QNMs of asymptotically AdS black holes play a crucial role in the holographic description of quark-gluon plasmas \cite{Kovtun:2004de, Luzum:2008cw}.

The
QNMs of Gauss-Bonnet and Lovelock black holes in asymptotically flat and de Sitter space-times have been studied numerically in Refs. \cite{Konoplya:2004xx, Cuyubamba:2016cug, Yoshida:2015vua, Chen:2016fuy, Konoplya:2017lhs} and in \cite{Kuang:2017cgt} at the non-zero axion field. Exact solutions for 
five-dimensional Gauss- Bonnet-AdS black holes with the fixed coupling constant $\alpha = R^2/2$, with $R$  being the AdS radius, 
were founded in Refs. \cite{Gonzalez:2017gwa, Gonzalez:2010vv}. The gravitational QNMs and the dual hydrodynamic regime for Gauss-Bonnet-anti-de Sitter black holes with the planar horizon were analyzed in \cite{Grozdanov:2016fkt, Grozdanov:2016vgg}, and it was found that the QNMs of asymptotically AdS black holes in theories with higher curvature corrections may help to describe the regimen of intermediate t' Hooft coupling in the dual field theory.
Furthermore, it have been shown in \cite{Konoplya:2017ymp, Gonzalez:2017gwa, Grozdanov:2016fkt, Grozdanov:2016vgg}  that the characteristic quasinormal spectra of a spherical Einstein--Gauss--Bonnet-AdS black hole are qualitatively different from their Einsteinian analogues, because there are nonperturbative modes in the Gauss--Bonnet coupling constant $\alpha$. Moreover, the black holes present gravitational instabilities,  or eikonal instability,  for the new branch  of nonperturbative modes. This occurs also for the asymptotically flat space-times \cite{Dotti:2005sq, Takahashi:2010ye}, and in some parameter ranges the new, nonperturbative in $\alpha$ modes may be dominant, leading to the qualitatively different profile of gravitational ringdown. The new nonperturbative modes were found for several quite different situations: not only for the Gauss--Bonnet, but also for the fourth-order in curvature ($R^4$) theory \cite{Grozdanov:2016fkt, Grozdanov:2016fkt}, for asymptotically flat \cite{Gleiser:2005ra, Dotti:2005sq, Takahashi:2010ye} and AdS black holes \cite{Konoplya:2017ymp, Gonzalez:2017gwa, Grozdanov:2016fkt, Grozdanov:2016fkt} and black branes, for gravitational \cite{Konoplya:2017ymp, Grozdanov:2016fkt, Grozdanov:2016fkt} and scalar field perturbations \cite{Gonzalez:2017gwa}. On the other hand, it has been shown that four-dimensional Einstein-dilaton-Gauss-Bonnet theory admits a wormhole solution without introducing any exotic matter  known as Kanti-Kleihaus-Kunz wormhole \cite{Kanti:2011jz}, which is unstable against small perturbations for any values of its parameters. 
This kind of instability is driven by the purely imaginary mode, which is nonperturbative in the Gauss-Bonnet coupling  \cite{Cuyubamba:2018jdl}.

Therefore, the phenomena of nonperturbative modes seems to be quite general and independent
of the asymptotic behavior of a black hole (flat, dS, AdS), topology of the event horizon (black hole or black
brane), spin of the field under consideration (scalar, gravitational, etc.), and, possibly, even of the particular form of the higher curvature corrections (Gauss-Bonnet, Lovelock, $R^4$, etc.). So, this nonperturbative modes may produce astrophysical
noticeable imprints on the ringdown of four dimensional black holes  with higher curvature corrections \cite{Konoplya:2017zwo}, such as Einstein-Gauss-Bonnet-dilaton, Einstein-Weyl \cite{Dotti:2005sq, Takahashi:2010ye} and others. From the gauge/gravity duality point of view, these
modes lead to the eikonal instability of Gauss-Bonnet black holes at some critical values of coupling constants \cite{Konoplya:2017ymp, Gonzalez:2017gwa}, so that they determine possible constraints on the holographic applicability of the black hole backgrounds.

In this work we extend the previously  mentioned studies of the scalar test field to the fermionic test field to analyze the stability of its propagation in the background of a Einstein-Gauss-Bonnet-AdS black hole and to show the existence of nonperturbative in $\alpha$ fermionic modes. In order to reach this goal, we use the improved asymptotic iteration method (AIM) \cite{Cho:2009cj}, which is an improved version of the AIM proposed in Refs. \cite{Ciftci,Ciftci:2005xn}. 
Also, we employ the spectral method \cite{Boyd} by using the Chebyshev basis
when the improved AIM turns out to be difficult to apply effectively, which occurs for high overtone modes; this method has been applied in Refs. \cite{Finazzo:2016psx, Gonzalez:2017shu}. However, at a fixed Gauss-Bonnet coupling $\alpha =R^2/2$ we find an exact solution for the QNMs for fermionic fields in this background. It is known that there is exact solution of the master perturbation equations for this case. Despite it suffering from the eikonal instability \cite{Konoplya:2017ymp}, and thereby from the absence of convergence in $\ell$, the fixed $\ell$-perturbations have no such  problem.

Regarding some studies of fermionic fields in black hole backgrounds, 
Chandrasekhar studied the general properties of a massive Dirac field equation in the Kerr background \cite{Chandrasekhar1}, and it was recently shown that there  are infinitely long-lived QNMs (quasiresonances) in the fermions quasinormal spectrum \cite{Konoplya:2017tvu} for this spacetime. Also, massless and massive Dirac QNFs in the Schwarzschild black hole spacetime was studied  by using the third-order WKB method \cite{Cho:2003qe}, and by using continued fraction and Hill-determinant approaches. Moreover, the study was extended to the Schwarzschild-de Sitter black hole by using the sixth-order WKB method \cite{Zhidenko:2003wq}, and the Dirac QNMs of the Reissner-Nordstr\"om de Sitter black hole was studied by using the P\"oshl-Teller potential approximation \cite{Jing:2003wq}. Other interesting studies have been conducted in \cite{Catalan:2013eza, Gonzalez:2017ptj} for Lifshitz black holes, in \cite{Gonzalez:2014voa} for new type black hole in three space-time dimensions, in \cite{Becar:2014jia} for a two-dimensional charged dilatonic black hole and in \cite{Becar:2013qba} for Chern-Simons and BTZ black holes with torsion.

The paper is organized as follows: In Sec. II we give a brief review 
of the Einstein-Gauss-Bonnet-AdS black hole backgrounds. 
Sec. III gives exact and numerical data on quasinormal modes for a fermionic field test  at various values of $\alpha$ in the region of stability. Finally, in Sec. IV,  we present the conclusions.

\section{Einstein--Gauss--Bonnet--AdS black holes}
\label{Background}

The Lagrangian of the $D$-dimensional Einstein--Gauss--Bonnet theory with a cosmological constant $\Lambda$ is given by
\begin{equation}\label{gbg3}
  \mathcal{L}=-2\Lambda+R+\frac{\alpha}{2}(R_{\mu\nu\lambda\sigma}R^{\mu\nu\lambda\sigma}-4\,R_{\mu\nu}R^{\mu\nu}+R^2)~,
\end{equation}
where $\alpha$ is the Gauss--Bonnet coupling constant. For $D=4$, the Gauss--Bonnet term is a topological invariant. Thus, the Gauss--Bonnet term contributes only to $D > 4$ space-time dimensions. 
An exact black hole solution of the field equations 
of theory (\ref{gbg3}) is described by the following static spherically symmetric metric \cite{Boulware:1985wk}:
\begin{equation}\label{gbg4}
 ds^2=-f(r)dt^2+\frac{1}{f(r)}dr^2 + r^2\,d\Omega_n^2~,
\end{equation}
where $d\Omega_n^2$ is a $(n=D-2)$-dimensional line element
of $S^\kappa_n$ manifold of constant curvature $\kappa = \pm 1, 0$ and
\begin{equation}\label{fdef}
f(r)=\kappa-r^2\,\psi(r) \,,
\end{equation}
where $\psi(r)$ satisfies 
\begin{equation}\label{Wdef}
W[\psi]\equiv\frac{n}{2}\psi(1 + \aa\psi) - \frac{\Lambda}{n + 1} = \frac{\mu}{r^{n + 1}}\,,
\end{equation}
being $\mu$ an integration constant, proportional to mass.
Here the Gauss--Bonnet coupling constant $\aa$ is
$$\aa \equiv \alpha\frac{(n - 1) (n - 2)}{2}\,.$$
It is worth noting that for 
\begin{equation}\label{psidef}
  \psi(r)=\frac{4\left(\frac{\mu}{r^{n+1}}+\frac{\Lambda}{n+1}\right)}{n+\sqrt{n^2+8\aa n\left(\frac{\mu}{r^{n+1}}+\frac{\Lambda}{n+1}\right)}}\,,
\end{equation}
the black hole solution of (\ref{Wdef}), which goes over into the known Tangherlini solutions \cite{Tangherlini:1963bw}, allowing for a non-zero $\Lambda$-term. In this paper we shall consider just this branch of solutions, because it has the known Einsteinian ($\alpha =0$) asymptotically flat, de Sitter and anti-de Sitter limits. When $\Lambda =0$, there is another branch of asymptotically anti-de Sitter solutions, which do not have the Einsteinian limit. In the following, we fix $\kappa = 1$ in order to consider 
a compact (spherical) black hole with the event horizon radius $r_ H$. Also, in order to measure all quantities in units of length for any value of $D$, it is convenient
to express $\mu$ as 
 \cite{Cuyubamba:2016cug}
\begin{equation}\label{massdef}
  \mu=\frac{n\,r_H^{n-1}}{2}\left(1+\frac{\aa}{r_H^2}-\frac{2\Lambda  r_H^2}{n(n+1)}\right)\,.
\end{equation}
and
$\Lambda$ as
\begin{equation}\label{AdSlambda}
  \Lambda=-\frac{n(n+1)}{2R^2}\left(1-\frac{\aa}{R^2}\right)\,,
\end{equation}
where 
the AdS radius $R$ has been defined by the relation $\psi(r\rightarrow\infty)=-1/R^{2}$. 
The above equation implies that $\aa<R^2$. It is worth mentioning that, when $R^2/2 < \aa < R2$, the solution (\ref{psidef}) does not satisfy the relation $\psi(r\rightarrow\infty)=-1/R^{2}$ and describes a black hole, which is identical to the one with $\aa < R^2/2$ after some re-scaling of parameters. Therefore, in the following we will consider black holes with $\aa < R^2/2$ \cite{Konoplya:2017zwo}. In $D=5$ case $\alpha =\aa$ and when, in addition, $\alpha = R^2/2 $, the metric function $f(r)$ has a BTZ-like form
\begin{equation}\label{metric}
f(r)=\frac{r^2}{R^2}+1-\sqrt{\frac{4\mu}{3R^2}}=\frac{r^2-r_H^2}{R^2}\,.
\end{equation}
This simplification of the metric function will make it possible allow to obtain an analytical solution to the Dirac equation, as we will show in the next section.

\section{Fermionic quasinormal modes} 
A minimally coupled fermionic field to curvature in the background of a
Einstein--Gauss--Bonnet--AdS black holes is given by the Dirac equation in curved space
\begin{equation}
\label{DE}
\left( \gamma ^{\mu }\nabla _{\mu }+m\right) \psi =0~,
\end{equation}%
where the covariant derivative is defined as 
$\nabla _{\mu }=\partial _{\mu }+\frac{1}{2}\omega _{\text{ \ }\mu
}^{ab}J_{ab}$,
with  $\omega^{ab}$ being the Levi-Civita spin connection and $J_{ab}=\frac{1}{4}\left[ \gamma _{a},\gamma _{b}\right]$, the generators of the Lorentz group are defined through the gamma matrices in a flat spacetime $\gamma ^{a}$, which can be expressed through 
the gamma matrices in curved space-time $\gamma ^{\mu }$ by  
$\gamma ^{\mu }=e_{\text{ \ }a}^{\mu }\gamma ^{a}$.
In order to
solve the Dirac equation (\ref{DE}), we use the diagonal vielbein
\begin{equation}
e^{0} = \sqrt{f} \, dt, \, \, \, e^{1} = \frac{1}{\sqrt{f}}dr,\, \, \, e^{2} = r \, d\theta, \, \, \, e^{3} = r \sin \theta \, d\phi, \, \, \, e^{4} = r \sin \theta \sin \phi \, d \varphi\,,
\end{equation}
and from the null torsion condition 
$de^{a}+\omega _{\text{ \ }b}^{a}\wedge e^{b}=0$,
the nonzero components of the spin connection are given by
\begin{align}
    \omega ^{01} &= \frac{f'(r)}{2} \, dt, \,\, \omega ^{12} = -\sqrt{f} \, d\theta, \,\, \omega ^{13} = -\sqrt{f} \sin \theta \,d\phi, \,\, \omega ^{14} = -\sqrt{f} \sin \theta \sin \phi \, d\varphi, \nonumber \\
    \omega ^{23} &= -\cos \theta \, d\phi, \,\, \omega ^{24} = -\cos \theta \sin \phi d\varphi, \,\,  \omega ^{34} = -\cos \phi \, d\varphi\,.
\end{align}
Now, using the following representation of the gamma matrices 
$\gamma ^{0}=i\sigma ^{2}\otimes \mathbf{1}~,\text{ \ }\gamma ^{1}=\sigma
^{1}\otimes \mathbf{1}~,\text{ \ }\gamma ^{m}=\sigma ^{3}\otimes \tilde{%
\gamma}^{m}$,
where $\sigma ^{i}$ are the Pauli matrices, and $\tilde{\gamma}^{m}$ are the
Dirac matrices in the base manifold $\Omega_n$, along with the
following ansatz for the fermionic field 
\begin{equation}
\psi =\frac{e^{-i\omega t}}{r^{n/2}f^{1/4} }\left( 
\begin{array}{c}
\psi _{1} \\ 
\psi _{2}%
\end{array}%
\right) \otimes \varsigma ~,
\end{equation}%
where $\varsigma $ is a $n$-components fermion. The following equations are
thus obtained:
\begin{eqnarray}
\label{diff}
\notag -\frac{i\omega}{\sqrt{f}}\psi _{2}+\sqrt{f}\psi'_{2}+\frac{\lambda}{r}\psi _{1}+m\psi _{1} &=&0~,   \\ 
\frac{i\omega}{\sqrt{f}}\psi _{1}+\sqrt{f}\psi'_{1}-\frac{\lambda}{r}\psi _{2}+m\psi _{2} &=&0~,  
\end{eqnarray}%
where $\lambda$ is the eigenvalue of the Dirac operator on the $n$-dimensional sphere and  is given by $\lambda= \pm i (\ell+n/2)$, with $\ell=0,1,2, ...$ and the prime denotes the derivative with respect to the radial coordinate $r$. These equations can be decoupled as
\begin{equation} \label{q1}
\psi_1''+\left(\frac{1}{2}\frac{f'(r)}{ f(r)}+\frac{\lambda }{r (\lambda -m r)}\right)\psi_1'+\frac{r^2 \omega  (m r-\lambda ) \left(2 \omega -i f'(r)\right)-2 f(r) \left((\lambda +m r) (\lambda -m r)^2+i \lambda  r \omega \right)}{2 r^2 f(r)^2 (m r-\lambda )} \psi_1=0\,,
\end{equation}
\begin{equation} \label{q2}
\psi_2''+\left(\frac{1}{2}\frac{f'(r)}{ f(r)}+\frac{\lambda }{r (\lambda +m r)}\right)\psi_2'+\frac{r^2 \omega  (m r+\lambda ) \left(2\omega +i f'(r)\right)-2 f(r) \left((-\lambda +m r) (\lambda +m r)^2+i \lambda  r \omega \right)}{2 r^2 f(r)^2 (m r+\lambda )} \psi_2=0\,.
\end{equation}
Notice that (\ref{q2}) can be obtained from (\ref{q1}) by means of the substitutions $\psi_1 \rightarrow \psi_2$, $\omega \rightarrow -\omega$ and $\lambda \rightarrow -\lambda$.

\subsection{Stability of massless fermionic field}

The above equations (\ref{diff}), for a massless fermionic field, can be reduced to \begin{eqnarray}
\label{a1}&& -i \omega Z_{+}-\frac{dZ_{-}}{d r^{\ast}}= W Z_{-}~,  \\
\label{a2}&& -i \omega Z_{-}-\frac{dZ_{+}}{dr^{\ast}}= -W Z_{+}~,
\end{eqnarray}
where we have defined $Z_{\pm}=\psi_1 \pm i \psi_2$ and $W=-i \lambda \sqrt{f}/r$, see \cite{Chandra}, and the tortoise coordinate $r^{\ast}$ is defined as usual by $d r^{\ast}=dr/f$. Now, decoupling (\ref{a1}) and (\ref{a2}), we obtain the following Schr\"odinger-like equations:
\begin{equation} \label{Schrodinger}
-\frac{d^2Z_\pm}{d r^{\ast 2}}+V_\pm=\omega^2 Z_\pm~, 
\end{equation}
where the effective potentials $V_{\pm}$ are given by
\begin{equation}
V_\pm=W^2 \pm \frac{dW}{d r^{\ast}} = \pm i \lambda \frac{f \sqrt{f}}{r^2} \mp i\lambda \frac{f' \sqrt{f}}{2r}-\frac{\lambda^2 f}{r^2}~.
\end{equation}
We can observe that the potentials are not positive-definite.
Suitable boundary conditions for the fermionic field are a purely ingoing wave at the event horizon and it vanishes at spatial infinity:
\begin{equation}\label{bc}
Z_{\pm}\big|_{r^{\ast} \rightarrow -\infty} \propto e^{-i \omega r^{\ast}}~, \,\,\,\,\, Z_{\pm}\big|_{r^{\ast} \rightarrow 0} \rightarrow 0~.
\end{equation}
Now, we shall show that the classical stability of the massless fermionic field in this background can be proven using the S-deformation method \cite{Kodama:2003ck}. So, multiplying Eq. (\ref{Schrodinger}) by $Z_{\pm}^*$, then performing an integration by parts and taking into account the boundary conditions (\ref{bc}), one can arrive at the following expression:
\begin{equation}\label{condition}
\int_{-\infty}^{0} \left( V_{\pm}(r) |Z_{\pm}(r^{\ast})|^2+\left| \frac{dZ_{\pm}(r^{\ast})}{dr^{\ast}}\right|^2 \right) dr^{\ast}= Z_{\pm}^{\ast} \frac{d Z_{\pm}}{d r^{\ast}} \Big|_{r^{\ast}=0}+ i \omega |Z_{\pm}(r^{\ast}=-\infty)|^2+\omega^2 \int_{-\infty}^{0} |Z_{\pm}(r^{\ast})|^2 dr^{\ast}~.
\end{equation}
The first term on the right-hand side of this equation is zero by the boundary condition (\ref{bc}), and one may conclude that the imaginary part of $\omega$ is always negative when $V_{\pm}(r)>0$ in the region outside the event horizon, see \cite{Kodama:2003ck, Zhidenko:2009zx} for details. The potentials $V_{\pm}$ are not positive-definite; however, in the S-deformation method a new derivative $D=\frac{d}{dr^{\ast}}+S(r^{\ast})$ is defined, and the first integral in (\ref{condition}) can be written as
\begin{equation}
\int_{-\infty}^{0} \left( V_{\pm}(r) |Z_{\pm}(r^{\ast})|^2+\left| \frac{dZ_{\pm}(r^{\ast})}{dr^{\ast}}\right|^2 \right) dr^{\ast}=\int_{-\infty}^{0}\left( \tilde{V}_{\pm} |Z_{\pm}(r^{\ast})|^2+|D Z_{\pm}|^2 \right) dr^{\ast}-S(r^{\ast}) |Z_{\pm}(r^{\ast})|^2 \Bigg| _{r^{\ast}=-\infty}^{r^{\ast}=0}~,
\end{equation}
where $\tilde{V}_{\pm}=V_{\pm}+\frac{dS}{d r^{\ast}}-S^2$. Appropriate functions are given by $S=-W$ for $V_{+}$ and $S=W$ for $V_{-}$ \cite{Zhidenko:2009zx, LopezOrtega:2012hx}. In this case, $\tilde{V}_{\pm}=0$ and we have $W(r^{\ast}=-\infty) = 0$ and $W(r^{\ast}=0)$ is a constant, which ensures that the integral is positive, and thus guarantees the stability of the fermionic field. Having demonstrated that the propagation of massless fermionic fields is stable in this background, in the following, we shall analyze two cases separately: the first is the special case in five space-time dimensions with $\alpha=R^2/2$, where we obtain the analytical solution to the radial equation, and the second case corresponds to the general case $\alpha\neq R^2/2$ and $D=5,6$, where we obtain the QNFs numerically by using the improved AIM approach for $D=5$ and the shooting method for $D=6$.

\subsection{Analytical solution}
In this section we obtain analytically the eigenvalues of the wave equation in $D=5$ case and $\alpha=R^2/2$. In this case, as we have mentioned, the metric function reduces to the BTZ-like form $f(r)=\left(r^2-r_H^2\right)/R^2$, and the following substitutions 
\begin{equation}
\psi _{1}\pm \psi _{2}=\left( \cosh (\rho/2) \pm \sinh (\rho/2) \right) \left( \phi
_{1}\pm \phi _{2}\right)~,
\end{equation}
in Eqs. (\ref{diff}) and the change of variables $z=\tanh
^{2}\rho$, with $r = r_H \cosh{\rho}$, enable us to obtain the following equations 
\begin{align}
    2 \sqrt{z} (1 - z) \, \partial _{z}\phi_{1} + \left(\frac{i \omega R^{2}}{r_H} \, z^{-1/2}  - \frac{\lambda R}{r_H} \, z^{1/2} \right) \phi_{1} + \left(\frac{1}{2} + mR + \frac{i \omega R^{2}}{r_H} - \frac{\lambda R}{r_H} \right) \, \phi_{2} = 0\,,  \nonumber \\
    2 \sqrt{z} (1 - z) \, \partial _{z}\phi_{2} - \left(\frac{i \omega R^{2}}{r_H} \, z^{-1/2}  - \frac{\lambda R}{r_H} \, z^{1/2} \right) \phi_{2} + \left(\frac{1}{2} + mR - \frac{i \omega R^{2}}{r_H} + \frac{\lambda R}{r_{H}} \right) \, \phi_{1} = 0  \,.
\label{system}
\end{align}
So, by decoupling $\phi_1$ from this system of equations we obtain
\begin{eqnarray}
 && \phi''_{1}(z) + \frac{1 - 3z}{2z(1-z)} \phi' _{1}(z) 
   -  \frac{1}{16 r_H^{2} z^{2}(1-z)^{2}}\Big((r_H + 2 m r_H R)^{2} z + 4 r_H R (1 - z) (z \lambda + i R \omega) \nonumber \\  && -4 R^{2} (1 - z) (z \lambda^{2} + R^{2} \omega^{2})\Big) \phi_{1}(z) = 0\,,
\end{eqnarray}
and a similar equation for $\phi_2$ but with the replacement $\lambda \rightarrow -\lambda$ and $\omega \rightarrow -\omega$. Then, defining
\begin{equation}
\phi _{1}\left( z\right) =z^{\alpha }\left( 1-z\right) ^{\beta }F\left(
z\right) ~,
\end{equation}
with 
\begin{equation}
\alpha =- \frac{i R^{2} \omega}{2 r_H}\,, \,\,\, \beta =\frac{1}{4} (1 + 2mR)~,
\end{equation}%
the radial equation reduces to the Gauss's hypergeometric equation for $F\left( z\right) $
\begin{equation}
z\left( 1-z\right) F^{\prime \prime }\left( z\right) +\left( c-\left(
1+a+b\right) z\right) F^{\prime }\left( z\right) -abF\left( z\right) =0~,
\end{equation}
which has three regular singular points at $z=0$, $z=1$ and $z=\infty $. The solution is given by 
\begin{equation}
\phi _{1}=C_{1}z^{\alpha }\left( 1-z\right) ^{\beta }{_{2}F_{1}}\left(
a,b,c,z\right) +C_{2}z^{1/2-\alpha }\left( 1-z\right) ^{\beta }{_{2}F_{1}}%
\left( a-c+1,b-c+1,2-c,z\right) ~,
\end{equation}
where, $_{2}F_{1}(a,b,c;z)$ denotes the hypergeometric function and $C_{1}$, $C_{2}$
are integration constants, with the other constants being defined as
\begin{eqnarray}
 \notag a &=& \frac{1}{2} + \alpha + \beta - \frac{ \lambda R}{2 r_H}\,,  \\
 \notag b &=& \alpha + \beta + \frac{ \lambda R}{2 r_H}~\,, \\
c&=&\frac{1}{2} + 2\alpha ~.
\end{eqnarray}%
Now, by imposing as boundary conditions at the horizon that there are only
ingoing modes implies that $C_{2}=0$. Thus, the solution simplifies to 
\begin{equation}
\phi _{1}\left( z\right) =C_{1}z^{\alpha }\left( 1-z\right) ^{\beta }{%
_{2}F_{1}}\left( a,b,c,z\right) ~.
\end{equation}
On the other hand, using Kummer's formula for hypergeometric functions \cite{M. Abramowitz}, 
\begin{eqnarray}
{_{2}F_{1}}\left( a,b,c,z\right) &=&\frac{\Gamma \left( c\right) \Gamma
\left( c-a-b\right) }{\Gamma \left( c-a\right) \Gamma \left( c-b\right) }{%
_{2}F_{1}}\left( a,b,a+b-c,1-z\right) + \notag  \label{form}
 \\
&&\left( 1-x\right) ^{c-a-b}\frac{\Gamma \left( c\right) \Gamma \left(
a+b-c\right) }{\Gamma \left( a\right) \Gamma \left( b\right) }{_{2}F_{1}}%
\left( c-a,c-b,c-a-b+1,1-z\right)~,
\end{eqnarray}
the behavior of the field at the sptial infinity $\left( z\rightarrow 1\right) $ is
given by 
\begin{equation}\label{A}
\phi _{1}\left( z\rightarrow 1\right) =C_{1}\left( 1-z\right) ^{\beta }\frac{%
\Gamma (c)\Gamma (c-a-b)}{\Gamma (c-a)\Gamma (c-b)}+C_{1}(1-z)^{-\beta }%
\frac{\Gamma (c)\Gamma (a+b-c)}{\Gamma (a)\Gamma (b)}~.
\end{equation}
Imposing that the fermionic field vanishes at spatial infinity yields the condition
 $a=-n$ or $b=-n$, where $n=0,1,2,...$. Therefore, the following sets of QNFs are obtained:
\begin{eqnarray}
 \notag \omega_1 R & = & \pm(\ell+\frac{3}{2})-\frac{i}{2} \frac{r_H}{R}(4n+3+2mR)\,, \\
\omega_2 R& = & \pm(\ell+\frac{3}{2})-\frac{i}{2}\frac{r_H}{ R}(4n+1+2mR)\,,
\end{eqnarray}
which can be combined into one set as
\begin{equation}
\omega R=\pm (\ell+\frac{3}{2})-\frac{i}{2}\frac{r_H}{R}(2n+1+2mR)\,.
\end{equation}

QNFs have real and imaginary parts, and the imaginary part is negative, which guarantees that the propagation of the fermionic field is stable in this background. Similarly, decoupling $\phi_2$ from the system of equations (\ref{system}), another set of quasinormal frequencies can be obtained. However, our numerical investigation shows that the same QNFs are obtained.

\subsection{Numerical solutions}
In this section, we obtain the QNFs numerically for a range of values of $\alpha$ and $D=5, 6$. We will employ the improved AIM for $D=5$, which is an improved version of the AIM  proposed in Refs. \cite{Ciftci, Ciftci:2005xn}, and the spectral method in the $D=6$ case. For $D=6$ we find that the improved AIM is difficult to apply effectively for high overtone modes.

In order to implement the improved AIM, we must consider the fermionic field and its behaviors at the horizon and at spatial infinity. It is convenient to introduce the coordinate $y=1-r_H/r$ So, at the horizon, $y\rightarrow 0$, the behavior of the fermionic
field is given by the solution of Eq. (\ref{q1}) at the horizon, which yields
\begin{equation}
\psi _{1}\left( y\rightarrow 0\right) \sim C_{1}y^{-\frac{i\omega r_H}{f'(0)}}+C_{2}y^{\frac{1}{2}+\frac{i\omega r_H}{f'(0)}}~,
\end{equation}
where, in the above equation and in the following the metric function is viewed as a function of the new radial coordinate $f(y)$, and the prime denotes derivative with respect to $y$.
So, in order to have only ingoing waves at the horizon, we impose $C_{2}=0$. 
On the other hand, at spatial infinity, from Eq. (\ref{q1}) we find that the fermionic field behaves as
\begin{equation}
\psi_{1} \left( y\rightarrow 1\right) \sim D_{1}\left( 1-y\right)
^A+D_{2}\left( 1-y\right) ^{-A}~,
\end{equation}
where $A=mR$ due to $\aa < R^2/2$. So, in order to have a null fermionic field at infinity we impose $D_{2}=0 $. 
Therefore, taking into account these behaviors we define the variable $\chi$ by 
\begin{equation}
\psi _{1}\left( y\right) =y^{-\frac{i\omega r_H}{f'(0)}}\left( 1-y\right) ^A\chi \left(
y\right)~,
\end{equation}
Then, by inserting this field into the radial equation, we obtain the homogeneous linear second-order differential equation for the function $\chi (y)$ 
\begin{equation}
\chi ^{\prime \prime }=\lambda _{0}(y)\chi ^{\prime }+s_{0}(y)\chi ~,
\label{de}
\end{equation}%
where
\begin{eqnarray}
\nonumber \lambda_0 (y)&=& \frac{-2 \left(2 m r_H \left(-i r_H (-1+y) \omega +(1+A) y f'(0)\right)+(-1+y) \lambda  \left(-2 i r_H (-1+y) \omega +(1+2 A) y f'(0)\right)\right)}{2 (-1+y) y (m r_H+(-1+y) \lambda ) f'(0)}\\
&& -\frac{f'(y)}{2 f(y) }~,
\end{eqnarray}
and
\begin{eqnarray}
\notag s_0(y) &=& \frac{r_H^2 \omega ^2}{y^2 f'(0)^2}-\frac{r_H^2 \omega ^2}{(y-1)^4 f(y)^2}+\frac{i r_H \omega  (m r_H (2 A y+y+1)+\lambda  (y-1) (2 A y+1))}{(y-1) y^2 f'(0) (m r_H+\lambda  (y-1))} \\
\notag && +f'(y) \left(\frac{\frac{A}{2-2 y}+\frac{i r_H \omega }{2 y f'(0)}}{f(y)}+\frac{i r_H \omega }{2 (y-1)^2 f(y)^2}\right) -\frac{A ((A+1) m r_H+A \lambda  (y-1))}{(y-1)^2 (m r_H+\lambda  (y-1))} \\
&& +\frac{(m r_H+\lambda  (y-1)) (\lambda +m r_H-\lambda  y)}{f(y)(y-1)^4}+\frac{i \lambda  r_H \omega }{f(y)(y-1)^2 (m r_H+\lambda  (y-1))} \,.
\end{eqnarray}
As we have mentioned, for $\psi _{2}$ our numerical solutions show the same QNFs that for $\psi _{1}$.

\begin{figure}[!h]
\begin{center}
\includegraphics[width=80mm]{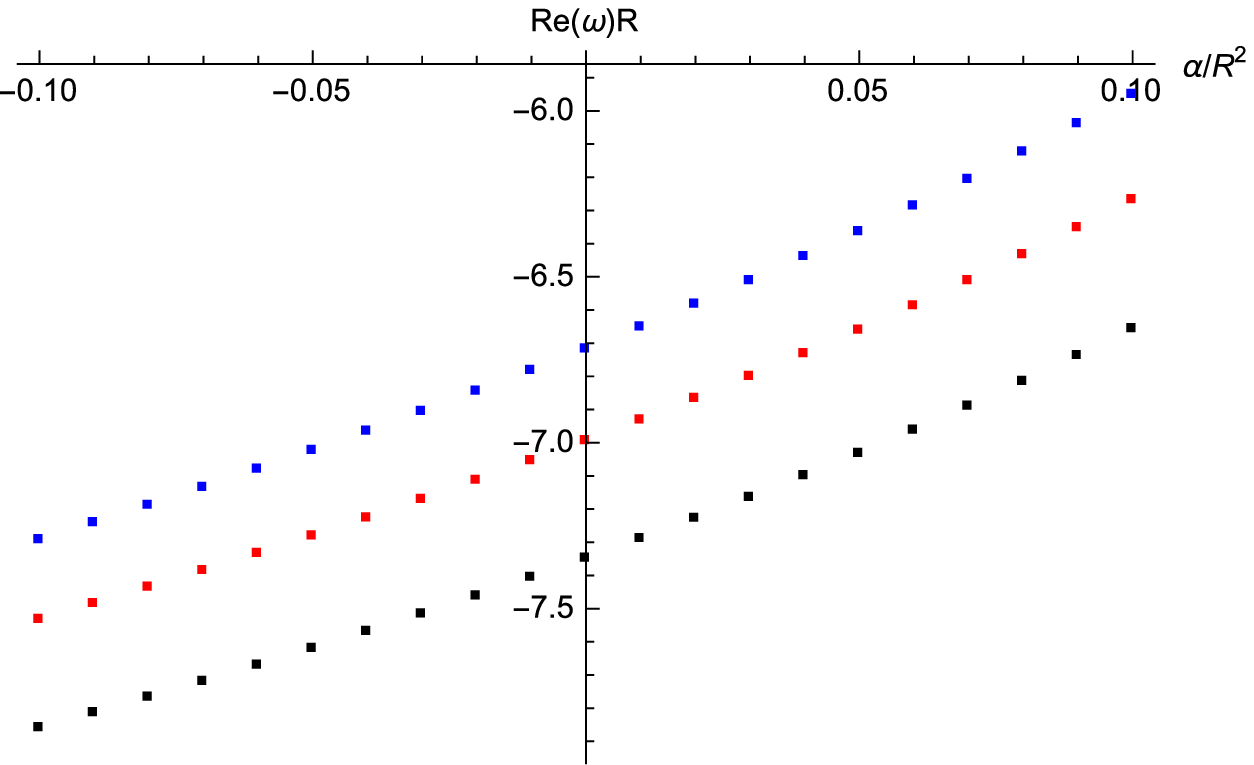}
\includegraphics[width=80mm]{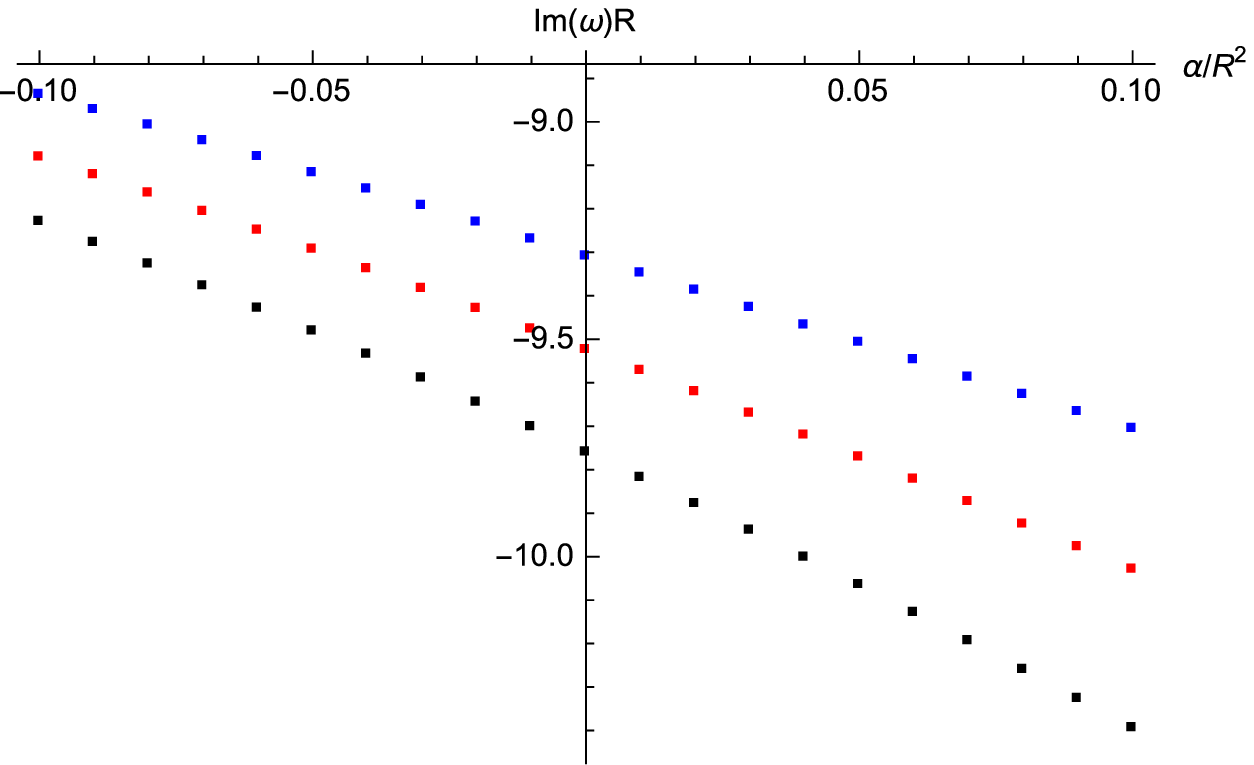}
\end{center}
\caption{The behavior of $Re(\omega)R$ (left panel) and $Im(\omega)R$ (right panel) versus $\alpha/R^2$ for the  perturbative branch in $\alpha$ with $D=5$, $r_H/R=5$, $\lambda = 1.5 i$ and different values of the mass $m$ and $\alpha$. $m R=0.01$ (blue, upper), $m R=0.1$ (red, middle) and $m R=0.2$ (black, lower)}
\label{FR}
\end{figure}

\begin{figure}[!h]
\begin{center}
\includegraphics[width=80mm]{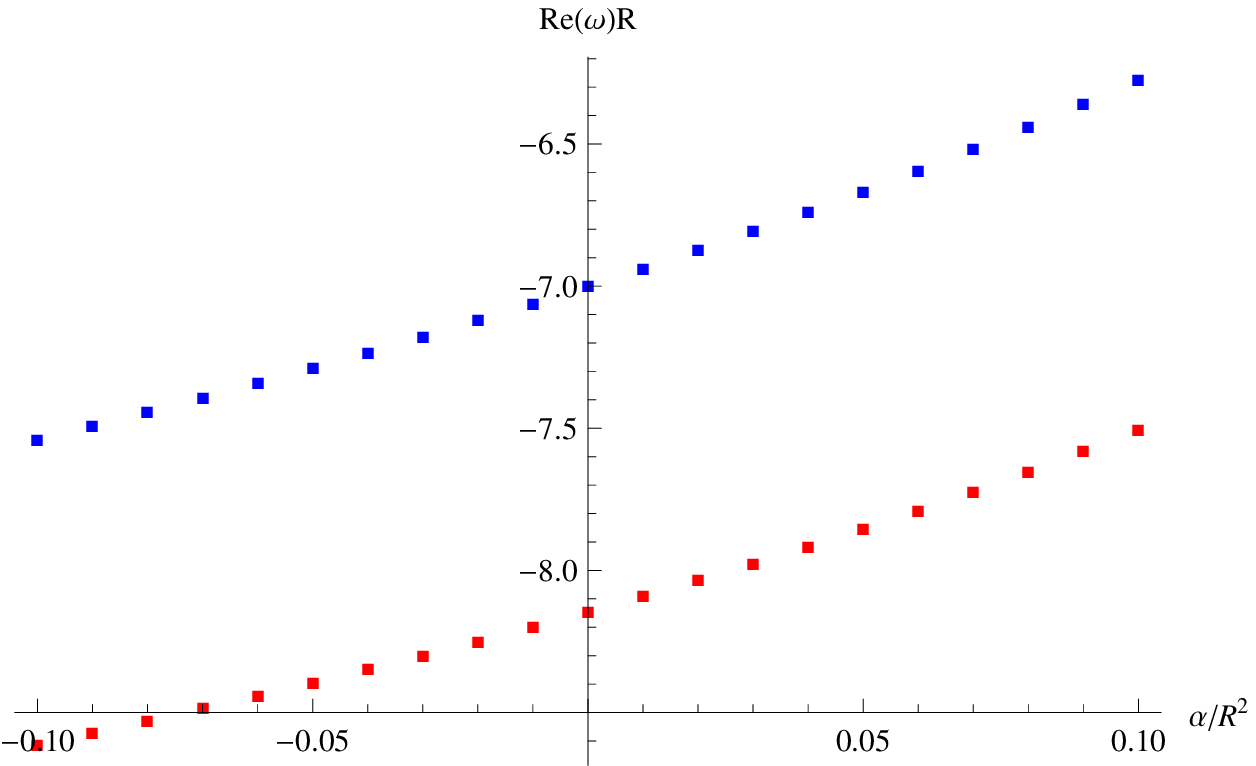}
\includegraphics[width=80mm]{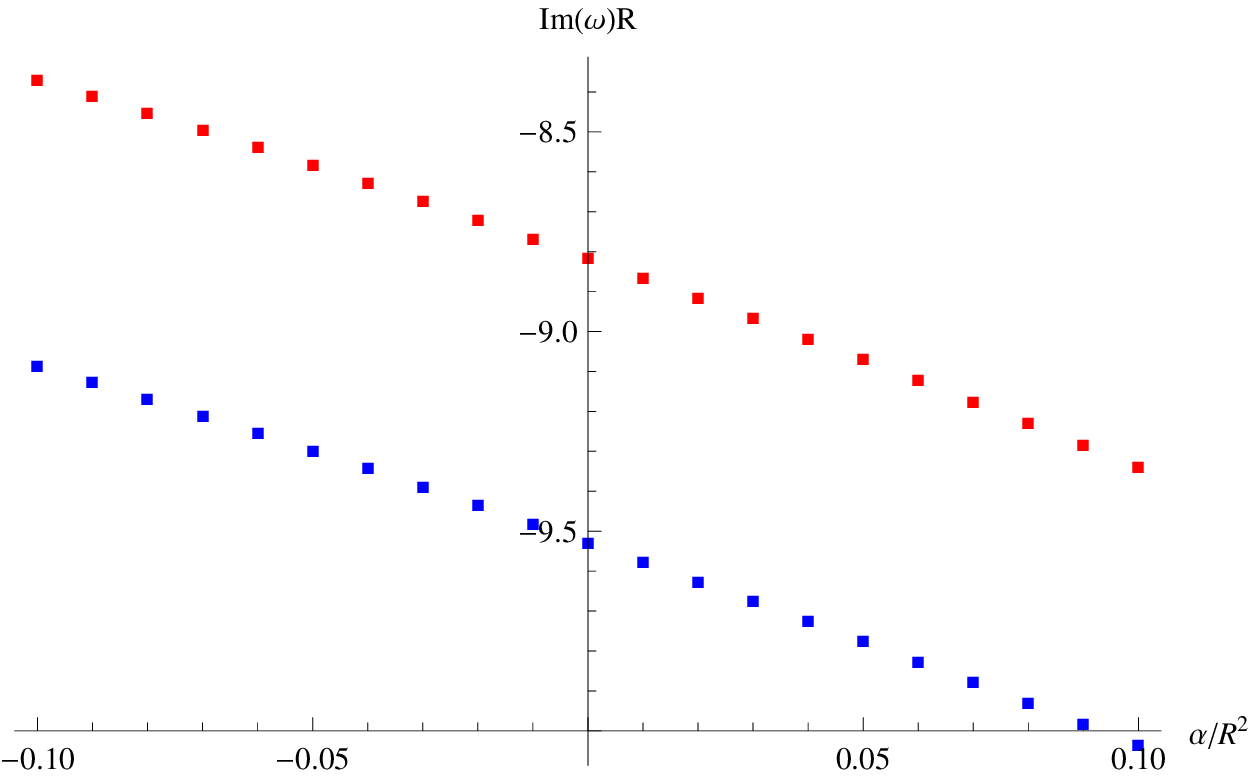}
\end{center}
\caption{The behavior of $Re(\omega)R$ (left panel) and $Im(\omega)R$ (right panel) versus $\alpha/R^2$ for the  perturbative branch in $\alpha$ with $D=5$, $r_H/R=5$ and $m R=0.1$. Blue points for $\lambda=1.5i$ (lower) and red points for $\lambda=2.5i$ (upper)}
\label{3PR}
\end{figure}

\begin{figure}[!h]
\begin{center}
\includegraphics[width=80mm]{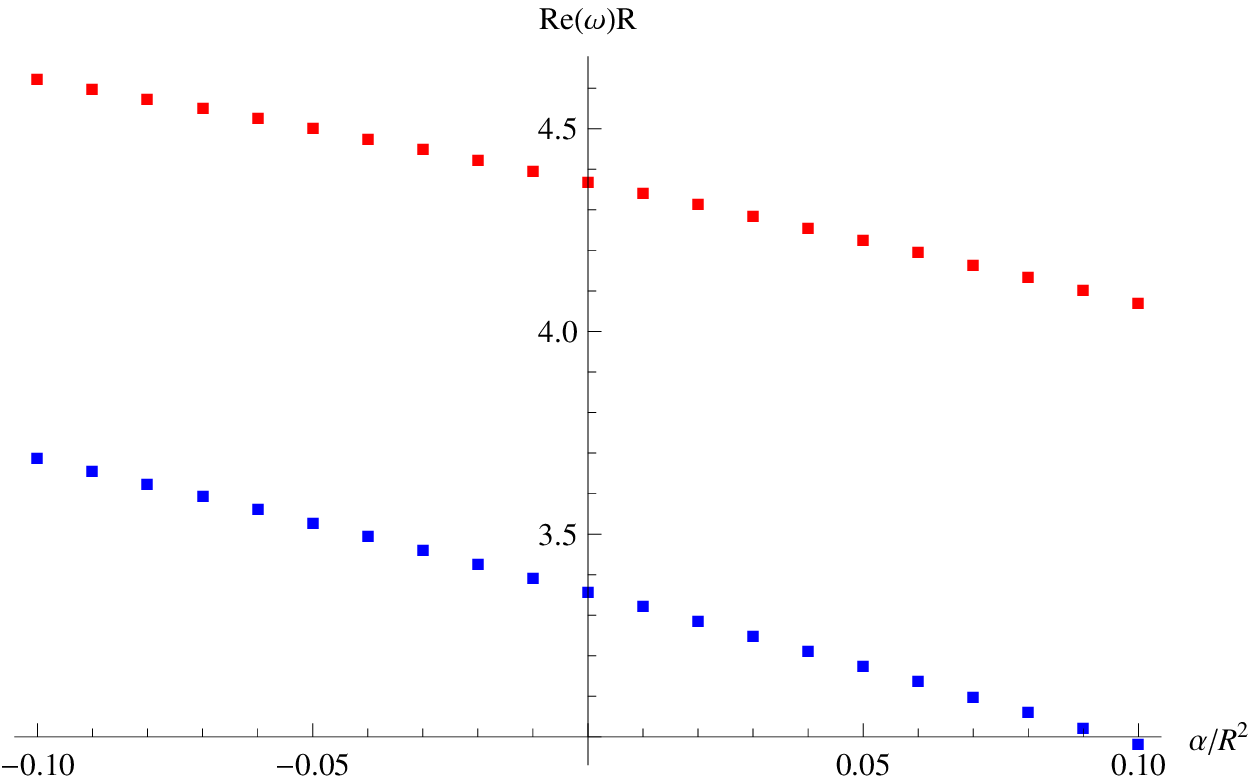}
\includegraphics[width=80mm]{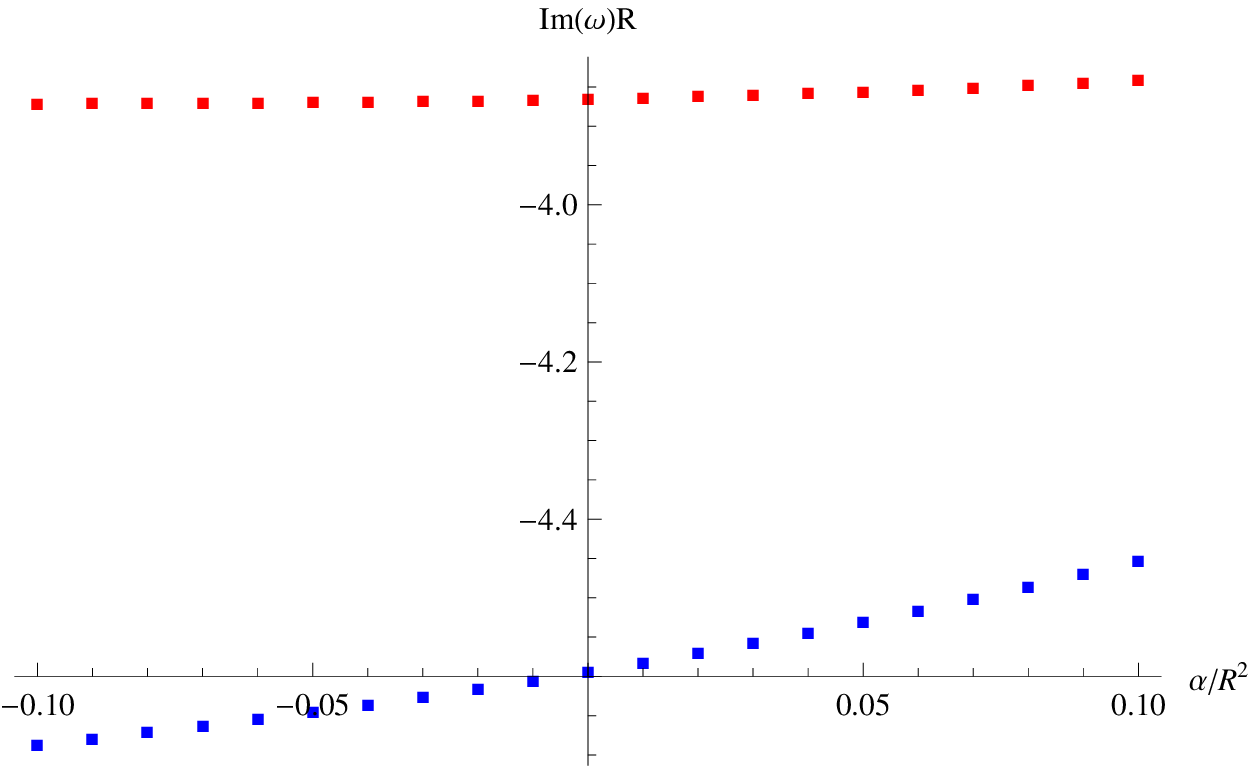}
\end{center}
\caption{The behavior of $Re(\omega)R$ (left panel) and $Im(\omega)R$ (right panel) versus $\alpha/R^2$ for the  perturbative branch in $\alpha$ with $D=5$, $r_H/R=5$ and $m R=0.1$. Blue points for $\lambda=1.5i$ (lower) and red points for $\lambda=2.5i$ (upper)}
\label{FNPR}
\end{figure}

Thus, in order to implement the improved AIM, it is necessary to differentiate Eq. (\ref{de}) $n$ times with respect to $y$,
which yields the following equation: 
\begin{equation}
\chi ^{n+2}=\lambda _{n}(y)\chi ^{\prime }+s_{n}(y)\chi~,  \label{de1}
\end{equation}%
where 
\begin{equation}
\lambda _{n}(y)=\lambda _{n-1}^{\prime }(y)+s_{n-1}(y)+\lambda
_{0}(y)\lambda _{n-1}(y)~,  \label{Ln}
\end{equation}%
and
\begin{equation}
s_{n}(y)=s_{n-1}^{\prime }(y)+s_{0}(y)\lambda _{n-1}(y)~.  \label{Snn}
\end{equation}%
Then, by expanding the $\lambda _{n}$ and $s_{n}$ in a Taylor series around
the point, $\xi $, at which the improved AIM is performed 
\begin{equation}
\lambda _{n}(\xi )=\sum_{i=0}^{\infty }c_{n}^{i}(y-\xi )^{i}~,
\end{equation}%
\begin{equation}
s_{n}(\xi )=\sum_{i=0}^{\infty }d_{n}^{i}(y-\xi )^{i}~,
\end{equation}%
where the $c_{n}^{i}$ and $d_{n}^{i}$ are the $i^{th}$ Taylor coefficients
of $\lambda _{n}(\xi )$ and $s_{n}(\xi )$, respectively, and by replacing
the above expansion in Eqs. (\ref{Ln}) and (\ref{Snn}), the following
set of recursion relations for the coefficients is obtained:
\begin{equation}
c_{n}^{i}=(i+1)c_{n-1}^{i+1}+d_{n-1}^{i}+%
\sum_{k=0}^{i}c_{0}^{k}c_{n-1}^{i-k}~,
\end{equation}%
\begin{equation}
d_{n}^{i}=(i+1)d_{n-1}^{i+1}+\sum_{k=0}^{i}d_{0}^{k}c_{n-1}^{i-k}~.
\end{equation}%
In this manner, by considering Taylor expansions the improved AIM avoided the
derivatives that contain the AIM in \cite{Cho:2009cj, Cho:2011sf}, and
the quantization conditions, which is equivalent to imposing a termination to the number of iterations \cite{Barakat:2006ki}, which is given by 
\begin{equation}
d_{n}^{0}c_{n-1}^{0}-d_{n-1}^{0}c_{n}^{0}=0~.
\end{equation}
By solving this equation numerically we find the QNFs. In Fig. \ref{FR} we plot the behavior of $Re(\omega)R$
and $Im(\omega)R$ for the first overtone
as a function of $\alpha / R^2$ 
for the  perturbative branch in $\alpha$ and different values of the mass of the fermionic field $m R=0.01, 0.1, 0.2$. We observe that the QNFs have real and imaginary parts, with the imaginary part being negative, which guarantees the stability of the propagation of the fermionic field. Also, when $\alpha$ increases, the real part of the QNFs increase and the imaginary part decrease.
Additionally, when $m R$ increase the real and imaginary part of the QNFs decrease. In Table \ref{QNM1} (see Appendix) we show the numerical values of QNFs plotted in Fig. \ref{FR}. 
Also, when $\lambda$ increases the real part of the QNFs decrease and the imaginary part increase. In Table \ref{QNM3} (see Appendix) we show the numerical values of the QNFs plotted in Fig. \ref{3PR}. 
Then, in Fig. \ref{FNPR} we plot the behavior of $Re(\omega)R$ and $Im(\omega)R$ for the fundamental QNF as a function of $\alpha / R^2$ 
for the  perturbative branch in $\alpha$ and different values of $\lambda$.  We observe that the QNFs have real and imaginary parts, with the imaginary part being negative, which guarantees the stability of the propagation of the fermionic field in this background. Also, when $\alpha$ increases, the real part of the QNFs decreases and the imaginary part increases.
Additionally, when $\lambda$ increase the real and imaginary parts of the QNFs increase. In Table \ref{QNM2} (see Appendix) we show the numerical values of the QNFs plotted in Fig. \ref{FNPR}. 

On the other hand, we consider the case $D=6$, and we find the fundamental QNFs for different values of the parameter $\tilde{\alpha}/R^2$ by using the spectral  method for the perturbative and nonperturbative  branches in $\alpha$. Notice that the QNFs have real and imaginary parts, with the imaginary part being negative, which guarantees the stability of the propagation of the fermionic field in this background; however, there is a gravitational instability in the region $0.42 \lesssim \tilde{\alpha}/R^2 \lesssim 0.6 $ \cite{Konoplya:2017ymp}. Interestingly enough, in contrast to gravitational and scalar field perturbations, where the nonperturbative modes are purely imaginary, we observe that the QNFs for the nonperturbative modes acquire a real part.

\begin{figure}[!h]
\begin{center}
\includegraphics[width=80mm]{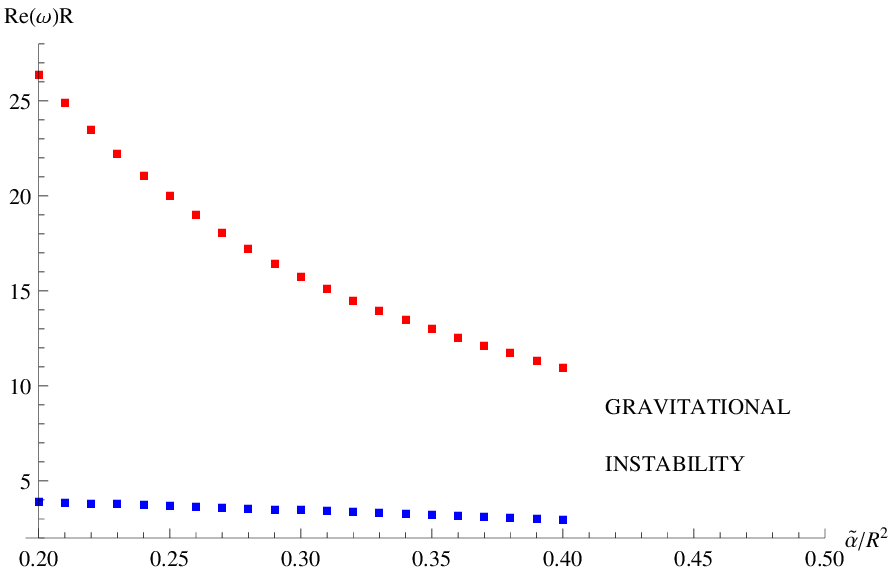}
\includegraphics[width=80mm]{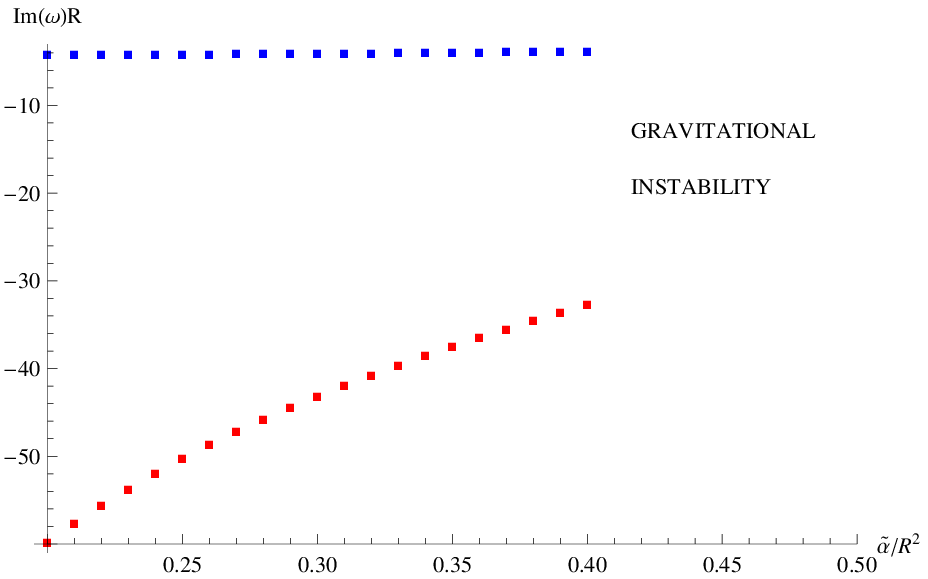}
\end{center}
\caption{The behavior of $Re(\omega)R$ (left panel) and $Im(\omega)R$ (right panel) for the perturbative  branch  in $\alpha$ (blue) and for the  nonperturbative branch in $\alpha$ (red) with $D=6$, $r_H/R=5$, $\lambda= 2 i$ and $mR =0.1$. The region $0.42 \lesssim \tilde{\alpha}/R^2 \lesssim 0.6 $  corresponds to (eikonal) gravitational instability \cite{Konoplya:2017ymp}.}
\label{FPRn4}
\end{figure}

\newpage

\section{Final remarks}
\label{conclusion}
In this work, we considered  Einstein--Gauss--Bonnet--AdS black holes as backgrounds, and we studied the propagation of a fermionic field in such backgrounds. We found the exact solution of the eigenvalues for the wave equation in $D=5$ space-time dimensions at the fixed value $\alpha=R^2/2$ of the Gauss--Bonnet coupling constant and numerical solutions for other values of the coupling constant $\alpha$ and $D=5, 6$ by using the improved AIM approach for $D=5$ and the spectral method for $D=6$. Mainly, we found two branches of QNFs, a perturbative branch  in the Gauss--Bonnet coupling constant $\alpha$, and another nonperturbative  branch  in $\alpha$, that is, they do not exist in the limit $\alpha=0$. The  nonperturbative branch in $\alpha$ is characterized by the growth of the imaginary part when $\alpha$ is decreasing, while for the  perturbative branch in $\alpha$ the QNFs tend to the QNFs of Schwarzschild AdS when $\alpha \rightarrow 0$.
We showed that the phenomena of nonperturbative modes, which
have been obtained in the spectrum of gravitational field perturbations and scalar field perturbations in previous works, also arise for fermionic field perturbations and therefore the phenomena of nonperturbative modes seems to be independent of the spin of the field under consideration.
However, in contrast to gravitational and scalar field perturbations, where the nonperturbative modes are purely imaginary, we found that for fermionic field perturbations the nonperturbative modes acquire a real part. 
The QNFs have real and imaginary parts for both branches, and the imaginary part is always negative; therefore, the spherical Gauss--Bonnet--AdS black holes are stable against fermionic field perturbations.\\

\newpage

\appendix{{\bf Appendix: Numerical quasinormal frequencies}}
\bigskip

In this appendix we show tables with the numerical values of the QNFs plotted in some figures
\begin{table}[ht]
\caption{Improved AIM. Quasinormal frequencies of fermionic field for perturbative branch, with $D=5$, $r_H/R=5$, $\lambda = 1.5 i$ and different values of the mass of the femionic field $m$ and the coupling constant $\alpha$.}
\label{QNM1}\centering
\begin{tabular}{|c | c | c | c | c | c | c | c | c |}
\hline\hline
$mR$ & $\alpha/R^2$ & $Re(\omega)R $  & $Im(\omega)R $ & 
$\alpha/R^2$ & $Re(\omega)R $  & $Im(\omega)R $ \\[0.5ex] \hline \
${}$ & $0.1$ & $-5.95382$ & $-9.70816$ & $0$ & $-6.72108$ & $-9.31105$  \\
${}$ & $0.09$ & $-6.04212$ & $-9.66922$ & $-0.01$ & $-6.78574$ & $-9.27200$ \\
${}$ & $0.08$ & $-6.12752$ & $-9.62979$ & $-0.02$ & $-6.84861$ & $-9.23331$ \\
${}$ & $0.07$ & $-6.21014$ & $-9.59002$  & $-0.03$ & $-6.90976$ & $-9.19501$ \\
${0.01}$ & $0.06$ & $-6.29012$ & $-9.55004$ & $-0.04$ & $-6.96927$ & $-9.15712$   \\
${}$ & $0.05$ & $-6.29012$ & $-9.55004$  & $-0.05$ & $-7.02723$ & $-9.11968$ 
\\ 
${}$ & $0.04$ & $-6.44264$ & $-9.46988$  & $-0.06$ & $-7.08369$ & $-9.08269$ \\
${}$ & $0.03$ & $-6.51542$ & $-9.42988$ & $-0.07$ & $-7.13873$ & $-9.04617$  
\\ 
${}$ & $0.02$ & $-6.51542$ & $-9.42988$  & $-0.08$ & $-7.19240$ & $-9.01014$
 \\

${}$ & $0.01$ & $-6.65454$ & $-9.35041$ & $-0.09$ & $-7.24477$ & $-8.97460$
  \\ [0.5ex] \hline
 ${}$ & $0.1$ & $-6.27102$ & $-10.03184$ & $0$ & $-6.99740$ & $-9.52655$ \\
${}$ & $0.09$ & $-6.35544$ & $-9.97986$  & $-0.01$ & $-7.05811$ & $-9.47899$ \\
${}$ & $0.08$ & $-6.43683$ & $-9.92795$  & $-0.02$ & $-7.11710$ & $-9.43213$ \\
${}$ & $0.07$ & $-6.51536$ & $-9.87623$ & $-0.03$ & $-7.17446$ & $-9.38599$  \\
${0.1}$ & $0.06$ & $-6.59119$ & $-9.82481$ & $-0.04$ & $-7.23026$ & $-9.34056$ \\
${}$ & $0.05$ & $-6.66447$ & $-9.77377$  & $-0.05$ & $-7.28458$ & $-9.29587$  \\ 
${}$ & $0.04$ & $-6.73535$ & $-9.72319$  & $-0.06$ & $-7.33750$ & $-9.25191$ \\
${}$ & $0.03$ & $-6.80396$ & $-9.67313$ & $-0.07$ & $-7.38908$ & $-9.20869$ \\ 
${}$ & $0.02$ & $-6.87043$ & $-9.62365$  & $-0.08$ & $-7.43938$ & $-9.16621$
 \\
${}$ & $0.01$ & $-6.93488$ & $-9.57477$ & $-0.09$ & $-7.48846$ & $-9.12446$  \\ [0.5ex] \hline
  ${}$ & $0.1$ & $-6.65996$ & $-10.39631$ & $0$ & $-7.35192$ & $-9.76198$  \\
${}$ & $0.09$ & $-6.74101$ & $-10.32891$   & $-0.01$ & $-7.40944$ & $-9.70417$ \\
${}$ & $0.08$ & $-6.81893$ & $-10.26220$ & $-0.02$ & $-7.46531$ & $-9.64745$  \\
${}$ & $0.07$ & $-6.89394$ & $-10.19628$ & $-0.03$ & $-7.51963$ & $-9.59182$ \\
${0.2}$ & $0.06$ & $-6.96623$ & $-10.13123$ & $-0.04$ & $-7.57248$ & $-9.53727$  \\
${}$ & $0.05$ & $-7.03598$ & $-10.06713$ & $-0.05$ & $-7.62393$ & $-9.48379$
 \\ 
${}$ & $0.04$ & $-7.10335$ & $-10.00401$ & $-0.06$ & $-7.67406$ & $-9.43137$
 \\
${}$ & $0.03$ & $-7.16850$ & $-9.94191$ & $-0.07$ & $-7.72293$ & $-9.38000$ 
 \\ 
${}$ & $0.02$ & $-7.23157$ & $-9.88086$ & $-0.08$ & $-7.77060$ & $-9.32966$
 \\
${}$ & $0.01$ & $-7.29267$ & $-9.82088$ & $-0.09$ & $-7.81714$ & $-9.28034$ 
 \\ [0.5ex] \hline

\end{tabular}%
\end{table}
\begin{table}[ht]
\caption{Improved AIM. Quasinormal frequencies of fermionic field for perturbative branch, with $D=5$, $r_H/R=5$, $mR=0.1$
and different values of $\lambda$ and $\alpha$.}
\label{QNM3}\centering
\begin{tabular}{|c|c|c|c|c|c|c|c|c|}
\hline\hline
$\lambda$ & $\alpha/R^2$ & $Re(\omega)R $  & $Im(\omega)R $ & 
$\alpha$ & $Re(\omega)R $  & $Im(\omega)R $ \\ [0.5ex] \hline
${}$ & $0.1$ & $-6.27102$ & $-10.03184$  & $0$ & $-6.99740$ & $-9.52655$ \\
${}$ & $0.09$ & $-6.35544$ & $-9.97986$ & $-0.01$ & $-7.05811$ & $-9.47899$  \\
${}$ & $0.08$ & $-6.43683$ & $-9.92795$ & $-0.02$ & $-7.11710$ & $-9.43213$ \\
${}$ & $0.07$ & $-6.51536$ & $-9.87623$ & $-0.03$ & $-7.17446$ & $-9.38599$ \\
${1.5i}$ & $0.06$ & $-6.59119$ & $-9.82481$ & $-0.04$ & $-7.23026$ & $-9.34056$  \\
${}$ & $0.05$ & $-6.66447$ & $-9.77377$ & $-0.05$ & $-7.28458$ & $-9.29587$
 \\ 
${}$ & $0.04$ & $-6.73535$ & $-9.72319$ & $-0.06$ & $-7.33750$ & $-9.25191$
 \\
${}$ & $0.03$ & $-6.80396$ & $-9.67313$ & $-0.07$ & $-7.38908$ & $-9.20869$ 
 \\ 
${}$ & $0.02$ & $-6.87043$ & $-9.62365$ & $-0.08$ & $-7.43938$ & $-9.16621$
  \\
${}$ & $0.01$ & $-6.93488$ & $-9.57477$ & $-0.09$ & $-7.48846$ & $-9.12446$
  \\ [0.5ex] \hline

  ${}$ & $0.1$ & $-7.50293$ & $-9.33772$  & $0$ & $-8.14275$ & $-8.81410$  \\
${}$ & $0.09$ & $-7.57797$ & $-9.28251$ & $-0.01$ & $-8.19563$ & $-8.76583$ \\
${}$ & $0.08$ & $-7.65015$ & $-9.22779$ & $-0.02$ & $-8.246929$ & $-8.71836$  \\
${}$ & $0.07$ & $-7.71962$ & $-9.17363$ & $-0.03$ & $-8.29672$ & $-8.67169$ \\
${2.5i}$ & $0.06$ & $-7.78657$ & $-9.12010$ & $-0.04$ & $-8.34509$ & $-8.62583$ \\
${}$ & $0.05$ & $-7.85113$ & $-9.06724$ & $-0.05$ & $-8.39211$ & $-8.58077$ \\ 
${}$ & $0.04$ & $-7.91345$ & $-9.01509$  & $-0.06$ & $-8.43785$ & $-8.53649$ \\
${}$ & $0.03$ & $-7.97365$ & $-8.96369$ & $-0.07$ & $-8.48237$ & $-8.49300$ \\ 
${}$ & $0.02$ & $-8.03186$ & $-8.91304$ & $-0.08$ & $-8.52572$ & $-8.45029$ \\
${}$ & $0.01$ & $-8.08819$ & $-8.86318$  & $-0.09$ & $-8.56798$ & $-8.40835$
 \\ [0.5ex] \hline

\end{tabular}%
\end{table}

\begin{table}[ht]
\caption{Improved AIM. Quasinormal frequencies of fermionic field for nonperturbative branch, with $D=5$, $r_H/R=5$, $mR=0.1$
and different values of $\lambda$ and $\alpha$.}
\label{QNM2}\centering
\begin{tabular}{|c|c|c|c|c|c|c|c|c|}
\hline\hline
$\lambda$ & $\alpha/R^2$ & $Re(\omega)R$  & $Im(\omega)R$ & 
$\alpha/R^2$ & $Re(\omega)R $  & $Im(\omega)R $ \\ [0.5ex] \hline
${}$ & $0.1$ & $2.98441$ & $-4.45140$  & $0$ & $3.36017$ & $-4.59301$  \\
${}$ & $0.09$ & $3.02395$ & $-4.46847$ & $-0.01$ & $3.39521$ & $-4.60410$  \\
${}$ & $0.08$ & $3.06309$ & $-4.48482$ & $-0.02$ & $3.42977$ & $-4.61472$  \\
${}$ & $0.07$ & $3.10180$ & $-4.50048$ & $-0.03$ & $3.46386$ & $-4.62491$ \\
${1.5i}$ & $0.06$ & $3.14008$ & $-4.51547$  & $-0.04$ & $3.49747$ & $-4.63468$ \\
${}$ & $0.05$ & $3.17791$ & $-4.52983$ & $-0.05$ & $3.53060$ & $-4.64404$ 
\\ 
${}$ & $0.04$ & $3.21530$ & $-4.54358$  & $-0.06$ & $3.56326$ & $-4.65303$
 \\
${}$ & $0.03$ & $3.25222$ & $-4.55675$ & $-0.07$ & $3.59546$ & $-4.66166$ 
\\ 
${}$ & $0.02$ & $3.28868$ & $-4.56936$  & $-0.08$ & $3.62718$ & $-4.66994$
  \\
${}$ & $0.01$ & $3.32466$ & $-4.58144$ & $-0.09$ & $3.65845$ & $-4.67790$
\\ [0.5ex] \hline

  ${}$ & $0.1$ & $4.07169$ & $-3.83977$  & $0$ & $4.37161$ & $-3.86384$ \\
${}$ & $0.09$ & $4.10406$ & $-3.84343$  & $-0.01$ & $4.39884$ & $-3.86504$  \\
${}$ & $0.08$ & $4.135891$ & $-3.84675$ & $-0.02$ & $4.425610$ & $-3.86609$ \\
${}$ & $0.07$ & $4.16717$ & $-3.84977$ & $-0.03$ & $4.45192$ & $-3.86699$ \\
$2.5i$ & $0.06$ & $4.19792$ & $-3.85250$ & $-0.04$ & $4.47778$ & $-3.86775$  \\
${}$ & $0.05$ & $4.22814$ & $-3.85496$  & $-0.05$ & $4.50321$ & $-3.86838$
\\ 
${}$ & $0.04$ & $4.25784$ & $-3.85717$  & $-0.06$ & $4.52820$ & $-3.86891$
 \\
${}$ & $0.03$ & $4.28703$ & $-3.85914$  & $-0.07$ & $4.55278$ & $-3.86932$
 \\ 
${}$ & $0.02$ & $4.31571$ & $-3.86090$ & $-0.08$ & $4.57694$ & $-3.86964$
  \\
${}$ & $0.01$ & $4.34390$ & $-3.86247$ & $-0.09$ & $4.60070$ & $-3.86987$
\\ [0.5ex] \hline

\end{tabular}%
\end{table}


\newpage 

\acknowledgments


This work was funded by the Direcci\'{o}n de Investigaci\'{o}n y Desarrollo de la Universidad de La Serena (Y. V., R. N. V.). P. A. G. acknowledges the hospitality of the Universidad de La Serena where part of this work was undertaken.


\newpage

\begin{thebibliography}{99}

\bibitem{Regge:1957td} 
  T.~Regge and J.~A.~Wheeler,
  Phys.\ Rev.\  {\bf 108}, 1063 (1957).
  
  
  
\bibitem{Zerilli:1971wd} 
  F.~J.~Zerilli,
  Phys.\ Rev.\ D {\bf 2}, 2141 (1970).
  
  
\bibitem{Kokkotas:1999bd} 
  K.~D.~Kokkotas and B.~G.~Schmidt,
  Living Rev.\ Rel.\  {\bf 2}, 2 (1999)
  [gr-qc/9909058].
  
  
\bibitem{Nollert:1999ji} 
  H.~P.~Nollert,
  Class.\ Quant.\ Grav.\  {\bf 16}, R159 (1999).
  
  
\bibitem{Konoplya:2011qq}
  R.~A.~Konoplya and A.~Zhidenko,
  Rev.\ Mod.\ Phys.\  {\bf 83}, 793 (2011)
  [arXiv:1102.4014 [gr-qc]];
  
  
  E.~Berti, V.~Cardoso and A.~O.~Starinets,
  Class.\ Quant.\ Grav.\  {\bf 26}, 163001 (2009)
  [arXiv:0905.2975 [gr-qc]];
  K.~D.~Kokkotas and B.~G.~Schmidt,
  Living Rev.\ Rel.\  {\bf 2}, 2 (1999)
  [gr-qc/9909058].


\bibitem{Abbott:2016blz}
  B.~P.~Abbott {\it et al.} [LIGO Scientific and Virgo Collaborations],
  Phys.\ Rev.\ Lett.\  {\bf 116}, no. 6, 061102 (2016)

\bibitem{TheLIGOScientific:2016src}
  B.~P.~Abbott {\it et al.} [LIGO Scientific and Virgo Collaborations],
  Phys.\ Rev.\ Lett.\  {\bf 116}, no. 22, 221101 (2016)

\bibitem{Konoplya:2016pmh}
  R.~Konoplya and A.~Zhidenko,
  Phys.\ Lett.\ B {\bf 756}, 350 (2016)

\bibitem{Kovtun:2004de}
  P.~Kovtun, D.~T.~Son and A.~O.~Starinets,
  Phys.\ Rev.\ Lett.\  {\bf 94}, 111601 (2005)

\bibitem{Luzum:2008cw}
  M.~Luzum and P.~Romatschke,
  Phys.\ Rev.\ C {\bf 78}, 034915 (2008)
  Erratum: [Phys.\ Rev.\ C {\bf 79}, 039903 (2009)]


\bibitem{Konoplya:2004xx}
  R.~Konoplya,
  Phys.\ Rev.\ D {\bf 71}, 024038 (2005)
  [hep-th/0410057];
  E.~Abdalla, R.~A.~Konoplya and C.~Molina,
  Phys.\ Rev.\ D {\bf 72}, 084006 (2005)


\bibitem{Cuyubamba:2016cug}
  M.~A.~Cuyubamba, R.~A.~Konoplya and A.~Zhidenko,
  Phys.\ Rev.\ D {\bf 93}, no. 10, 104053 (2016)

\bibitem{Yoshida:2015vua}
  D.~Yoshida and J.~Soda,
  Phys.\ Rev.\ D {\bf 93}, no. 4, 044024 (2016)

\bibitem{Chen:2016fuy}
  B.~Chen and P.~C.~Li,
  arXiv:1607.04713 [hep-th].



\bibitem{Konoplya:2017lhs}
  R.~A.~Konoplya and A.~Zhidenko,
  JCAP {\bf 1705}, no. 05, 050 (2017)
  [arXiv:1705.01656 [hep-th]].
  
  
\bibitem{Kuang:2017cgt}
  X.~M.~Kuang and J.~P.~Wu,
  arXiv:1702.01490 [hep-th].
  
  
\bibitem{Gonzalez:2010vv} 
  P.~Gonzalez, E.~Papantonopoulos and J.~Saavedra,
  JHEP {\bf 1008}, 050 (2010)
  [arXiv:1003.1381 [hep-th]].
  
  
  \bibitem{Gonzalez:2017gwa} 
  P.~A.~Gonzalez, R.~A.~Konoplya and Y.~Vasquez,
  Phys.\ Rev.\ D {\bf 95}, no. 12, 124012 (2017)
  [arXiv:1703.06215 [gr-qc]].


 \bibitem{Grozdanov:2016fkt}
  S.~Grozdanov and A.~O.~Starinets,
  arXiv:1611.07053 [hep-th].



 \bibitem{Grozdanov:2016vgg}
  S.~Grozdanov, N.~Kaplis and A.~O.~Starinets,
  JHEP {\bf 1607}, 151 (2016)
  [arXiv:1605.02173 [hep-th]].
 

\bibitem{Konoplya:2017ymp} 
  R.~A.~Konoplya and A.~Zhidenko,
  Phys.\ Rev.\ D {\bf 95}, no. 10, 104005 (2017)
  [arXiv:1701.01652 [hep-th]].



  
  
  
\bibitem{Takahashi:2010ye} 
  T.~Takahashi and J.~Soda,
  Prog.\ Theor.\ Phys.\  {\bf 124}, 911 (2010)
  [arXiv:1008.1385 [gr-qc]].


  





\bibitem{Dotti:2005sq}
  G.~Dotti, R.~J.~Gleiser,
  Phys.\ Rev.\ D {\bf 72}, 044018 (2005)
  [gr-qc/0503117].


\bibitem{Gleiser:2005ra}
  R.~J.~Gleiser, G.~Dotti,
  Phys.\ Rev.\ D {\bf 72}, 124002 (2005)
  [gr-qc/0510069].


  
\bibitem{Kanti:2011jz} 
  P.~Kanti, B.~Kleihaus and J.~Kunz,
  Phys.\ Rev.\ Lett.\  {\bf 107}, 271101 (2011)
  [arXiv:1108.3003 [gr-qc]].
  
  
   \bibitem{Cuyubamba:2018jdl}
  M.~A.~Cuyubamba, R.~A.~Konoplya and  A.~Zhidenko,
  arXiv:1804.11170 [gr-qc].
  
  \bibitem{Konoplya:2017zwo}
  R.~A.~Konoplya and A.~Zhidenko,
  JHEP {\bf 1709}, 139 (2017)
  [arXiv:1705.07732 [hep-th]].
  
  
\bibitem{Cho:2009cj} 
  H.~T.~Cho, A.~S.~Cornell, J.~Doukas and W.~Naylor,
  Class.\ Quant.\ Grav.\  {\bf 27}, 155004 (2010)
  [arXiv:0912.2740 [gr-qc]].
  
  

\bibitem{Ciftci} 
H. Ciftci, R. L. Hall, and N. Saad, J. Phys. A 36(47), 11807-11816 (2003).


\bibitem{Ciftci:2005xn} 
  H.~Ciftci, R.~L.~Hall and N.~Saad,
  Phys.\ Lett.\ A {\bf 340}, 388 (2005).
  

\bibitem{Boyd}
J. P. Boyd, Chebyshev and Fourier Spectral Methods. Dover Books on Mathematics. Dover Publications, Mineola, NY, second ed., 2001.
  
  
\bibitem{Finazzo:2016psx} 
  S.~I.~Finazzo, R.~Rougemont, M.~Zaniboni, R.~Critelli and J.~Noronha,
  JHEP {\bf 1701}, 137 (2017)
  [arXiv:1610.01519 [hep-th]].
  
\bibitem{Gonzalez:2017shu} 
  P.~A.~Gonzalez, E.~Papantonopoulos, J.~Saavedra and Y.~Vasquez,
  Phys.\ Rev.\ D {\bf 95}, no. 6, 064046 (2017)
  [arXiv:1702.00439 [gr-qc]].
  
  
  
  \bibitem{Chandrasekhar1}
 S. Chandrasekhar, Proc. Roy. Soc. Lond. A 349, 571
(1976).
  
\bibitem{Konoplya:2017tvu} 
  R.~A.~Konoplya and A.~Zhidenko,
  Phys.\ Rev.\ D {\bf 97}, no. 8, 084034 (2018)
  [arXiv:1712.06667 [gr-qc]].
  
  
\bibitem{Cho:2003qe} 
  H.~T.~Cho,
  Phys.\ Rev.\ D {\bf 68}, 024003 (2003)
  [gr-qc/0303078].
  

  
  
\bibitem{Zhidenko:2003wq} 
  A.~Zhidenko,
  Class.\ Quant.\ Grav.\  {\bf 21}, 273 (2004)
  [gr-qc/0307012].
  
  
 
  


\bibitem{Jing:2003wq} 
  J.~l.~Jing,
  Phys.\ Rev.\ D {\bf 69}, 084009 (2004)
  [gr-qc/0312079].
  
  
  
  
   \bibitem{Gonzalez:2017ptj} 
  P.~A.~Gonzalez, Y.~Vasquez and R.~N.~Villalobos,
  Eur.\ Phys.\ J.\ C {\bf 77}, no. 9, 579 (2017)
  [arXiv:1704.00413 [hep-th]].
  

\bibitem{Catalan:2013eza} 
  M.~Catalan, E.~Cisternas, P.~A.~Gonzalez and Y.~Vasquez,
  Eur.\ Phys.\ J.\ C {\bf 74}, no. 3, 2813 (2014)
  [arXiv:1312.6451 [gr-qc]].
  
  \bibitem{Gonzalez:2014voa} 
  P.~A.~Gonzalez and Y.~Vasquez,
  Eur.\ Phys.\ J.\ C {\bf 74}, no. 7, 2969 (2014)
  [arXiv:1404.5371 [gr-qc]].
  
  
  
\bibitem{Becar:2014jia} 
  R.~Becar, P.~A.~Gonzalez and Y.~Vasquez,
  Eur.\ Phys.\ J.\ C {\bf 74}, 2940 (2014)
  [arXiv:1405.1509 [gr-qc]].
  
  
  \bibitem{Becar:2013qba} 
  R.~Becar, P.~A.~Gonzalez and Y.~Vasquez,
  Phys.\ Rev.\ D {\bf 89}, no. 2, 023001 (2014)
  [arXiv:1306.5974 [gr-qc]].
  
  
  
  
  
  
  
  
  
 \bibitem{Boulware:1985wk}
  D.~G.~Boulware and S.~Deser,
 Phys.\ Rev.\ Lett.\  {\bf 55}, 2656 (1985).
   
  
 \bibitem{Tangherlini:1963bw}
  F.~R.~Tangherlini,
  Nuovo Cim.\  {\bf 27}, 636 (1963).
  
 
 \bibitem{Chandra}
Chandrasekhar S.:
The Mathematical Theory of Black Holes.
Oxford University Press, New York (1983).

  
  \bibitem{Kodama:2003ck} 
  H.~Kodama and A.~Ishibashi,
  gr-qc/0312012.
  
  
  \bibitem{Zhidenko:2009zx} 
  A.~Zhidenko,
  arXiv:0903.3555 [gr-qc].

\bibitem{LopezOrtega:2012hx} 
  A.~Lopez-Ortega,
  Int.\ J.\ Mod.\ Phys.\ D {\bf 21}, 1250092 (2012)
  [arXiv:1211.1801 [gr-qc]].
  
  
 \bibitem{M. Abramowitz}
 M. Abramowitz and A. Stegun, Handbook of
 Mathematical functions, (Dover publications, New York, 1970).


  
\bibitem{Cho:2011sf} 
  H.~T.~Cho, A.~S.~Cornell, J.~Doukas, T.~R.~Huang and W.~Naylor,
 Adv.\ Math.\ Phys.\  {\bf 2012}, 281705 (2012)
 [arXiv:1111.5024 [gr-qc]].
  
  
  
\bibitem{Barakat:2006ki}
  T.~Barakat,
  Int.\ J.\ Mod.\ Phys.\ A {\bf 21} (2006) 4127.
  

  

 






  
\end{thebibliography}
\end{document}